\documentstyle[aps,prd,floats,epsf]{revtex}
\begin{document}
\draft % Makes pacs numbers print

%%%%%%%%%%%%%%%%%%%%%%%%%%%%%%%
% Shortcuts
%%%%%%%%%%%%%%%%%%%%%%%%%%%%%%%

%%%%%%%%%%%%%%%%%%%%%%%%%%%%%%%

%--------------------------------------------------------------------
%--------------------------------------------------------------------
%  Two column format:
\twocolumn[\hsize\textwidth\columnwidth\hsize\csname
@twocolumnfalse\endcsname
%--------------------------------------------------------------------
%--------------------------------------------------------------------

\title{Gravitational Waves from the Dynamical Bar Instability
in a Rapidly Rotating Star}
\author{J. David Brown}
\address{Department of Physics\\
North Carolina State University\\
Raleigh, NC 27695--8202}
\date{\today}

\maketitle

\begin{abstract}
A rapidly rotating, axisymmetric star can be dynamically unstable to an $m=2$ ``bar" mode that transforms the star from a disk shape to an elongated bar. The fate of such a bar--shaped star is uncertain. 
Some previous numerical studies indicate that the bar is short lived, lasting for only a few bar--rotation periods, while other studies suggest that the bar  is relatively long lived. This paper contains the results of a numerical simulation of a rapidly rotating $\gamma=5/3$ fluid star. The simulation shows that the bar shape is long lived: once the bar is established, the star retains this shape for more than $10$ bar--rotation periods, through the end of the simulation. The results are consistent with the conjecture that a star will retain its bar shape indefinitely on a dynamical time scale, as long as its rotation rate exceeds the threshold for secular bar instability. The results are described in terms of a low density neutron star, but can be scaled to represent, for example, a burned--out stellar core that is prevented from complete collapse by centrifugal forces. Estimates for the gravitational--wave signal indicate that a dynamically unstable neutron star in our galaxy can be detected easily by the first generation of ground based gravitational--wave detectors. The signal for an unstable neutron star in the Virgo cluster might be seen by the planned advanced detectors. The Newtonian/quadrupole approximation is used throughout this work. 
\end{abstract}
% PACS numbers:

%--------------------------------------------------------------------
%--------------------------------------------------------------------
% Two column format: 
\vskip2pc]
%--------------------------------------------------------------------
%--------------------------------------------------------------------

\section{Introduction} \label{sec:intro}

A self--gravitating, axisymmetric fluid body with a sufficiently high rotation rate can be dynamically unstable to non-axisymmetric perturbations. Typically, the fastest growing unstable mode is the $m=2$ ``bar" mode which acts to transform the body from a disk--like shape
to an elongated bar that tumbles end--over--end. This instability has been described analytically for the case of uniform density bodies\cite{Chandra,ST}, and has been 
the subject of numerous numerical studies\footnote{For a brief review, 
see Ref.~\cite{NCT}}. The numerical results show that bar formation is 
accompanied by the ejection of mass and angular momentum, and that the ejected matter forms long spiral arms in the equatorial plane. The subsequent evolution is less certain. Some simulations indicate 
that the bar shape is short lived, with the star returning to a predominantly disk--like shape after a few bar--rotation periods. Other simulations predict that the bar  persists for many bar--rotation periods. In recent work, New, Centrella, and Tohline\cite{NCT} address this issue with a series of simulations using two different codes at various resolutions, and conclude that the bar shape is persistent. 
In their highest resolution run the bar decayed after roughly $6$ or $7$ bar--rotation periods. This was believed to be caused by numerical 
errors that induced an unphysical center of mass motion. In a lower resolution run a symmetry condition was imposed that prevented any center of mass motion. In that case the star maintained its bar shape throughout the simulation. 

The purpose of the present work is to simulate the long--time evolution of a rapidly rotating, self--gravitating star using Newtonian 
hydrodynamics and  gravity. As discussed in Sec.~\ref{sec:numcode}, the numerical code is substantially different from the codes that have been used previously to address this problem. The initial data for this study consists of a $\gamma = 5/3$ polytrope with stability parameter $\beta = 0.30$. The stability parameter is defined by $\beta = T/|W|$, where $T$ is rotational kinetic energy and $W$ is gravitational potential energy. The results here suggest that the bar shape is indeed long--lived---the star displays a prominent bar shape at the end of the simulation, which includes more than $10$ bar rotation periods. 

Numerical studies of fluids with various equations of state and initial rotation profiles have shown that the {\em dynamical} 
bar instability appears when the stability parameter $\beta$ exceeds a certain critical value, typically close to $0.27$\cite{DT,TIPD}. For the {\em secular} instability, which arises through dissipative mechanisms, the critical value of $\beta$ is near $0.14$. A neutron star might reach the critical value of $\beta$ for dynamical or secular 
instability by accreting matter and angular momentum from a binary companion. A stellar core that has exhausted its nuclear fuel might reach a critical rotation rate as it collapses. 

A star or stellar core that develops a rotating bar--like configuration will generate large amounts of gravitational radiation. 
Depending on the distance of the source, this radiation might be strong 
enough to be detected by the world--wide network of gravitational--wave detectors currently under construction\cite{BandC}. Here, the question of persistence of the bar shape becomes very important. The detectability of a source depends on its characteristic amplitude $h_c \approx h\sqrt{n}$, where $h$ is the amplitude of the waves
with frequency $f$ and $n$ is the number of wave cycles in a bandwidth near $f$\cite{Thorne}. Thus, long--duration signals with large $n$ can be more easily detected than short--duration signals. 

Should we expect the bar shape to persist or decay? One reason why the bar might decay is the loss of mass and angular momentum  from the ends of the bar. The accompanying drop in rotational kinetic energy could reduce the stability parameter and allow the star to return to 
axisymmetry. Loss of rotational kinetic energy through shock heating
might also occur. For the simulation presented here $\beta$ has an 
initial value of $0.30$, large enough to dynamically trigger the growth of the bar mode. During the initial period of bar formation, mass and 
angular momentum are shed from the ends of the bar and long, spiral arms are formed. The stability parameter rapidly drops below $0.27$ and eventually settles to a value of about $0.24$. However, these losses of mass and angular momentum, with the accompanying drop in $\beta$, 
are insufficient to completely rob the star of its bar shape. 

The results here are consistent with the conjecture that the star will 
retain its bar shape indefinitely on a dynamical time scale, as long as $\beta$ stays above the critical value for secular instability\cite{IDP}. If this is correct, the bar should 
still decay due to secular processes. Imamura, Durisen, and Pickett 
have argued that the bar continues to shed small amounts of angular momentum to the spiral arms, and these loses cause the bar to decay\cite{IDP}. Other secular mechanisms, such as energy and angular momentum loss through gravitational radiation, could play a role as well. 

It is possible that, for the equation of state and angular velocity profile used here, the star is able to retain its bar shape simply because the critical value of $\beta$ for dynamical instability is $0.24$ or less. This possibility was tested by conducting a 
second simulation using the same equation of state and scaling the angular velocity by a factor of $3/4$. The resulting model has an initial value of $\beta$ equal to $0.25$. The bar mode showed no signs of growth in this simulation. Thus, for the models tested here, the bar mode does not spontaneously grow unless $\beta$ exceeds a critical value greater than $0.25$. However, once the bar shape is established, it can persist for many bar--rotation periods with $\beta$ equal to $0.24$ or less. 

In the present numerical calculation there are two sources of error in angular momentum. First, some of the angular momentum that is lost off the edge of the computational grid might, ideally, fall back onto the star. Second, numerical errors at each time step introduce a purely artificial loss of angular momentum. Both of these errors act to decrease the angular momentum of the star. One expects such losses to cause the bar shape to deteriorate. What is seen, both in the gravitational wave analysis and in the Fourier analysis of the density, is a relatively persistent signal with a modest decline in amplitude over the duration of the simulation. 

The physical model and initial data that form the basis of the simulation are described in Sec.~II. Section III contains a discussion of the numerical code. The results of the simulation are described qualitatively and quantitatively in Sec.~IV. Conclusions and results are summarized in Sec.~V. Details of the initial data code are given in the Appendix. 

%%%%%%%%%%%%%%%%%%%%%%%%%%%%%%%%%%%%%%%%%%%%%%%%%%%%%%%%%%%%%%%%%%%%%%%%
\section{Physical model} \label{sec:physmod}

The initial data for the simulation consists of a stationary, axisymmetric Newtonian fluid star with polytropic equation of state, $P = K\rho^\gamma$. The equation of state parameters are chosen to coincide roughly with those of a low density neutron star with soft equation of state, $\gamma = 5/3$ and $K = 5.38\times 10^{9}\,{\rm cgs}$\cite{ST}. 
The bar instability in stellar cores with stiff equations of state has been investigated by Houser and Centrella\cite{HC}.

The equations of hydrostatic equilibrium are solved using an algorithm described in the Appendix, in which the freely specifiable data are the central density $\rho_{\rm c}$ and the angular velocity distribution $\omega(r)$. Here, $r$ is the distance from the rotation axis. With the choices 
\begin{mathletters}
\begin{eqnarray}
   \rho_{\rm c} & = & 2.00\times 10^{14} \,{\rm g}/{\rm cm}^3\ ,\\
   \omega(r)    & = &  (4000/{\rm s}) 
                       e^{-(r/4.80\times 10^6 \,{\rm cm})^2} \ ,
\end{eqnarray}
\end{mathletters}
the equilibrium configuration has mass, equatorial radius, polar radius, angular momentum, and stability parameter given by 
\begin{mathletters}
\begin{eqnarray}
   M          & = & 2.37\,M_\odot \ ,\\
   R_{\rm eq} & = & 4.91\times 10^6 \,{\rm cm} \ ,\\
   R_{\rm p}  & = & 1.11\times 10^6 \,{\rm cm} \ ,\\
   J          & = & 6.98\times 10^{49}\,{\rm g}\,{\rm cm}^2/{\rm s}\ ,\\
   \beta      & = & 0.300 \ ,
\end{eqnarray}
\end{mathletters}
respectively. Equations (1b) and (2b) show that the angular velocity 
has values 
\begin{equation}
   \omega(0) = 4000/{\rm s} \ ,\quad \omega(R_{\rm eq}) = 
    1400/{\rm s} \ , 
\end{equation}
on the rotation axis and at the equator. The azimuthal velocity at the 
equator is $6.89\times 10^9\,{\rm cm}/{\rm s}$, below the Kepler 
velocity of  $8.90\times 10^{9}\,{\rm cm}/{\rm s}$. Note that the mass (2a) is considerably larger than the masses of observed neutron stars, and exceeds the limit for nonrotating neutron stars for most equations of state. However, as discussed by Baumgarte, Shapiro, and Shibata\cite{BSS}, differentially rotating ``hypermassive" neutron stars with $M > 2\,M_\odot$ might appear as the immediate products of core--collapse supernovae or binary neutron star mergers. 

The rotation law for a uniform density Maclaurin spheroid is 
\begin{equation}
   j_{\rm sp}(m_{\rm en}) = (5J/2M)[1 - (1 - m_{\rm en}/M)^{2/3}] \ ,
\end{equation}
where $j_{\rm sp} = r^2 \omega$ is the specific angular momentum and $m_{\rm en}$ is the mass enclosed in a cylindrical radius. Many previous works on the dynamical bar instability have employed this same rotation law (see, for example, Refs.~\cite{BO,WT,HC,NCT}). For the initial data code used here the angular velocity is specified as a function of radius $r$, so the rotation law (4) cannot be used directly. However, the rotation law (1b) was chosen for its similarity to the Maclaurin law, as Fig.~1 shows.
%%%%%%%%%%%%%%%%%%%%%%%%%%%%%%%%%%%%%%%%%%%%%%%%%%%%%%%%
\begin{figure}[!htb]                                   %
% GNUPLOT: LaTeX picture with Postscript
\begingroup%
  \makeatletter%
  \newcommand{\GNUPLOTspecial}{%
    \@sanitize\catcode`\%=14\relax\special}%
  \setlength{\unitlength}{0.1bp}%
\begin{picture}(2448,1728)(0,0)%
\special{psfile=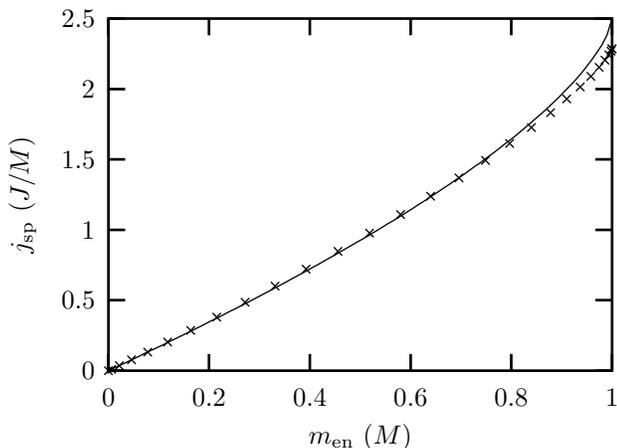 llx=0 lly=0 urx=490 ury=403 rwi=4900}
\put(1349,50){\makebox(0,0){$m_{\rm en}$ ($M$)}}%
\put(100,964){%
\special{ps: gsave currentpoint currentpoint translate
270 rotate neg exch neg exch translate}%
\makebox(0,0)[b]{\shortstack{$j_{\rm sp}$ ($J/M$)}}%
\special{ps: currentpoint grestore moveto}%
}%
\put(2298,200){\makebox(0,0){1}}%
\put(1918,200){\makebox(0,0){0.8}}%
\put(1539,200){\makebox(0,0){0.6}}%
\put(1159,200){\makebox(0,0){0.4}}%
\put(779,200){\makebox(0,0){0.2}}%
\put(400,200){\makebox(0,0){0}}%
\put(350,1628){\makebox(0,0)[r]{2.5}}%
\put(350,1362){\makebox(0,0)[r]{2}}%
\put(350,1097){\makebox(0,0)[r]{1.5}}%
\put(350,831){\makebox(0,0)[r]{1}}%
\put(350,566){\makebox(0,0)[r]{0.5}}%
\put(350,300){\makebox(0,0)[r]{0}}%
\end{picture}%
\endgroup
                                       % Figure 1
\caption{The specific angular momentum $j_{\rm sp}$,   %
in units of $J/M$, versus the mass $m_{\rm en}$        %
enclosed in a cylindrical radius, in units of $M$.     %
Both the numerical data (crosses $\times$) and the     %
Maclaurin rotation law (solid curve) are shown.}       %
\end{figure}                                           %
%%%%%%%%%%%%%%%%%%%%%%%%%%%%%%%%%%%%%%%%%%%%%%%%%%%%%%%%
In Fig.~1, the numerical data for $j_{\rm sp}$ and $m_{\rm en}$ are plotted for the rotation law (1b), along with the Maclaurin rotation law (4). For clarity, only one in five numerical data points is shown. The data and the rotation law (4) agree on the value of $j_{\rm sp}$ to within $2\%$ throughout most of the star, out to about $m_{\rm en}/M = 0.8$. At the surface of the star, the rotation law (1b) differs from the Maclaurin rotation law by about $8\%$. 

The dimensionful model described here can be rescaled by introducing 
parameters $\lambda$, $\mu$, and $\tau$ that scale the lengths, masses, 
and times. To be precise, a quantity $Q$ with dimensions 
$[Q] = L^A M^B T^C$, where $L$ is length, $M$ is mass, and $T$ is time, 
is rescaled according to $Q_{\rm new} = Q_{\rm old}/(\lambda^{A} \mu^{B} \tau^{C})$. A physical, dimensionful model is retained if the parameters are dimensionless and Newton's constant $G$ is unchanged by the rescaling. This requires $\lambda$, $\mu$, and $\tau$ to satisfy $\lambda^3 = \mu \tau^2$. Alternatively, the model can be converted to polytropic units\cite{WT} in which $G=K=M=1$ by setting  $\lambda = 4.81\times 10^5\,{\rm cm}$, $\mu = 4.71\times 10^{33}\,{\rm g}$, and 
$\tau = 1.88\times 10^{-5}\,{\rm s}$. This leads to 
\begin{mathletters}
\begin{eqnarray}
   \rho_{\rm c} & = & 4.73\times 10^{-3} \ ,\\
   R_{\rm eq}   & = & 10.2 \ ,\\
   \omega(0)    & = & 7.53\times 10^{-2} \ ,
\end{eqnarray}
\end{mathletters}
for the central density, equatorial radius, and central angular velocity. 

For the model with the scaling displayed in Eqs.~(1) and (2), the maximum flow velocity is $8.2\times 10^9\,{\rm cm}/{\rm s}$. This occurs at a distance of $3.4\times 10^6\,{\rm cm}$ from the rotation axis. At this peak velocity, the special relativistic gamma factor $(1 - v^2/c^2)^{-1/2}$, where $c$ is the speed of light, differs from unity by about 4\%. General relativistic effects should be greatest near the surface of the star at the poles, where the gravitational potential is approximately $GM/R_{\rm p} = 2.8\times 10^{20} \,{\rm cm}^2/{\rm s}^2$. Comparing this result with $c^2$, we find that general relativistic effects represent a correction of roughly 30\%  at the poles. Thus, with the scaling used in Eqs.~(1) and (2), the errors that we introduce by ignoring special and general relativity are substantial. 

The errors are reduced if the model is rescaled appropriately. For example, with $\lambda = 1/2$, $\mu = 2$, and $\tau = 1/4$, the 
mass of the star is cut in half, its linear size is increased by a factor of $2$, and its angular velocity is decreased by a factor of $4$. In this case the model might represent a stellar core that has partially collapsed, but is prevented from contracting to nuclear density by centrifugal forces. (Note that with this scaling Newton's constant is unchanged. Also note that the speed of light should not be scaled, since it does not appear in the Newtonian hydrodynamical equations or the Poisson equation.) The scaling reduces the maximum speed by a factor of 
$2$ and reduces the gravitational potential by a factor of $4$. For 
this model, special and general relativistic effects represent corrections of about 2\% and 8\%, respectively. 

%%%%%%%%%%%%%%%%%%%%%%%%%%%%%%%%%%%%%%%%%%%%%%%%%%%%%%%%%%%%%%%%%%%%%%%
\section{Numerical code} \label{sec:numcode}

The axisymmetric initial data was calculated on a grid in the $r$--$z$ 
plane consisting of  $512\,\times\, 511$ zones. Details and tests of the initial data code are contained in the Appendix. In preparation for the evolution of this data, the grid was chosen to cover a physical domain of $1.77\times 10^7\,{\rm cm}$ in radius and $2.50\times 10^7\,{\rm cm}$ in the $z$--direction. Thus, the star was contained in a small region of the grid, roughly the inner $142 \,\times\, 45$ zones. The quality of the data can be checked with the diagnostic expression ${\cal V} = (2T + W + 3\int P\,dV)/W$, where $T$ is kinetic energy, $W$ is gravitational potential energy, and $\int P\,dV$ is the volume integral of pressure. According to the virial equations, this quantity should vanish\cite{Chandra,Tassoul}. For the numerically generated model, this diagnostic has a value of ${\cal V} = 3.2\times 10^{-5}$. 

At the beginning of the evolution, the density and velocity are 
interpolated onto a three--dimensional Cartesian grid. The grid 
contains $128^3$ zones and covers a cubic domain with sides of length $2.50\times 10^7\,{\rm cm}$. The equatorial radius of the star spans $25$ zones, while the polar radius spans $6$ zones. The density in each zone is modified with a random perturbation ranging from $-10\%$ to $+10\%$ of the unperturbed value. Initially, the specific internal 
energy $e$ is obtained from the density by the relation $e = K\rho^{\gamma-1}/(\gamma-1)$. Thereafter, the star is evolved with the 
equation of state 
\begin{equation} 
   P = (\gamma-1)\rho e 
\end{equation}
with $\gamma = 5/3$. 

The evolution code includes the nonrelativistic hydrodynamics code VH-1, written by Blondin, Hawley, Lindahl and Lufkin. VH-1 is based on the piecewise--parabolic method (PPM) as described by Colella and Woodward\cite{CW}. The PPM scheme is a higher--order extension of Godunov's method\cite{Godunov}. It uses parabolas as interpolation functions within each zone, and characteristics to determine the domains of dependence for zone interfaces. The average values of density, pressure and velocity within these domains are used as inputs to the Riemann problem between adjacent zones. The solution of the 
Riemann problem determines time averaged values of pressure and velocity, which in turn are used to compute hydrodynamical fluxes. 
In VH-1, each time step is reduced to three one--dimensional evolutionary ``sweeps" via operator splitting. For each one--dimensional sweep, the fluid variables are evolved in Lagrangian coordinates and then remapped onto the original Eulerian grid. The order of the sweeps is cycled through all possible permutations. 

For the simulation presented here, I used a ``Courant number" of $0.3$. That is, the timestep was set to $0.3$ times the maximum allowed by the CFL condition. This is a relatively small timestep for a PPM code. It was chosen to help minimize the artificial loss of angular momentum, discussed below. 

The Poisson equation for the gravitational potential is solved using multigrid methods\cite{PTVF}. The finest grid has size $129^3$, with grid points that lie at the corners of the $128^3$ hydrodynamical zones. The boundary conditions are computed from a multipole expansion of the potential that includes monopole, dipole, quadrupole, and octupole terms. 

The PPM method is designed to evolve the fluid mass density $\rho$, 
linear momentum density, and total energy density, and in the process to conserve the mass, linear momentum, and total energy of the system. Unfortunately, for such a rapidly rotating star, the total energy density $E$ is dominated by kinetic energy density $K$. This is a problem because the internal energy density $\rho e = E - K$ is needed for the calculation of pressure from the equation of state (6). Since $\rho e$ is a difference of large numbers, it is subject to large numerical errors. As long as the fluid flow is smooth, one can solve this problem by modifying the code so that internal energy, rather than total energy, is evolved. However, this modification of the PPM algorithm leads to the wrong jump conditions at shock fronts. 
Thus, for the present simulation, a compromise was struck: in regions of smooth flow the code evolves internal energy, and in the neighborhood of shocks the code evolves total energy. Similar schemes have been described in Refs.~\cite{BNSCO,ZTFLOCDRRMT}.

VH-1 uses the shock flattening algorithm described by Colella and Woodward\cite{CW} as a dissipation mechanism. The algorithm determines a ``flattening coefficient" $f$ for each computational zone, which ranges from  $0$ for smooth flow to a maximum of $0.5$ in the presence of a  strong shock. The flattening coefficient is also used to determine whether internal energy or total energy is used to update the fluid variables in each zone. Thus, internal energy is used when $f$ is less than some threshold value, $f_t$, while total energy is used for $f > f_t$. In practice, the performance of the code was found to be insensitive to the value of the threshold $f_t$, for values ranging from $0.1$ to $0.5$. 

The ability of this ``hybrid" code to switch from internal energy to total energy at shock fronts was demonstrated on a variety of 
one--dimensional test problems. These include the standard Sod shock tube problem with density and pressure ratios of up to $1.0\times 10^6$, and strong standing shocks. Figures 2 and 3 show the results of the most extreme test, a standing shock wave with an upstream Mach number of $1.0\times 10^8$. Results are shown for the hybrid code, a total energy code (which evolves total energy exclusively) and an internal energy code (which evolves internal energy exclusively). 
%%%%%%%%%%%%%%%%%%%%%%%%%%%%%%%%%%%%%%%%%%%%%%%%%%%%%%%%
\begin{figure}[!htb]                                   %
% GNUPLOT: LaTeX picture with Postscript
\begingroup%
  \makeatletter%
  \newcommand{\GNUPLOTspecial}{%
    \@sanitize\catcode`\%=14\relax\special}%
  \setlength{\unitlength}{0.1bp}%
\begin{picture}(2448,1728)(0,0)%
\special{psfile=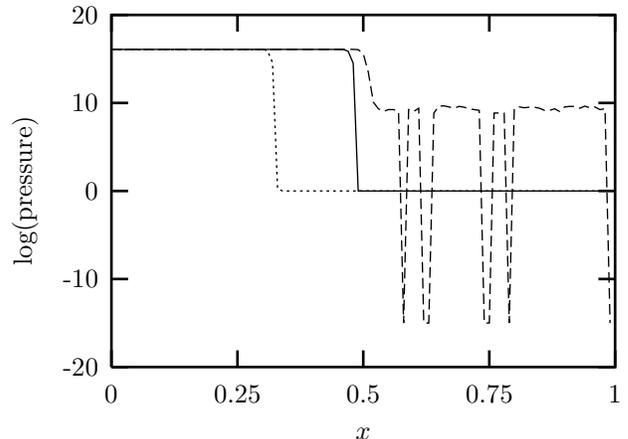 llx=0 lly=0 urx=490 ury=403 rwi=4900}
\put(1349,50){\makebox(0,0){$x$}}%
\put(100,964){%
\special{ps: gsave currentpoint currentpoint translate
270 rotate neg exch neg exch translate}%
\makebox(0,0)[b]{\shortstack{log(pressure)}}%
\special{ps: currentpoint grestore moveto}%
}%
\put(2298,200){\makebox(0,0){1}}%
\put(1824,200){\makebox(0,0){0.75}}%
\put(1349,200){\makebox(0,0){0.5}}%
\put(875,200){\makebox(0,0){0.25}}%
\put(400,200){\makebox(0,0){0}}%
\put(350,1628){\makebox(0,0)[r]{20}}%
\put(350,1296){\makebox(0,0)[r]{10}}%
\put(350,964){\makebox(0,0)[r]{0}}%
\put(350,632){\makebox(0,0)[r]{-10}}%
\put(350,300){\makebox(0,0)[r]{-20}}%
\end{picture}%
\endgroup
                                     % Figure 2
\caption{Logarithm of pressure for the strong standing %
shock test. The curves were obtained from the hybrid   %
code (solid), total energy code (dashed) and internal  %
energy code (dotted).}                                 %
\end{figure}                                           %
%%%%%%%%%%%%%%%%%%%%%%%%%%%%%%%%%%%%%%%%%%%%%%%%%%%%%%%%
%%%%%%%%%%%%%%%%%%%%%%%%%%%%%%%%%%%%%%%%%%%%%%%%%%%%%%%%
\begin{figure}[!htb]                                   %
% GNUPLOT: LaTeX picture with Postscript
\begingroup%
  \makeatletter%
  \newcommand{\GNUPLOTspecial}{%
    \@sanitize\catcode`\%=14\relax\special}%
  \setlength{\unitlength}{0.1bp}%
\begin{picture}(2448,1728)(0,0)%
\special{psfile=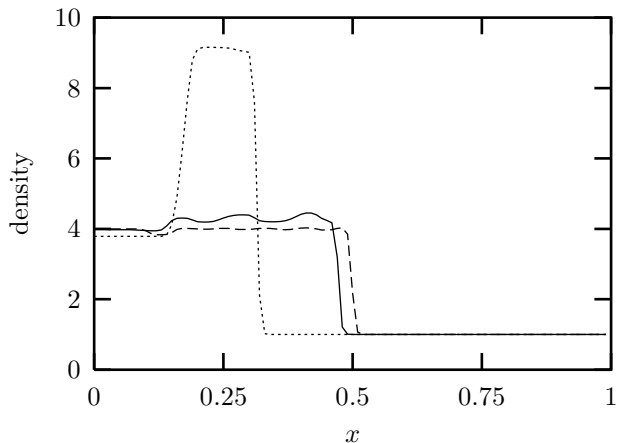 llx=0 lly=0 urx=490 ury=403 rwi=4900}
\put(1324,50){\makebox(0,0){$x$}}%
\put(100,964){%
\special{ps: gsave currentpoint currentpoint translate
270 rotate neg exch neg exch translate}%
\makebox(0,0)[b]{\shortstack{density}}%
\special{ps: currentpoint grestore moveto}%
}%
\put(2298,200){\makebox(0,0){1}}%
\put(1811,200){\makebox(0,0){0.75}}%
\put(1324,200){\makebox(0,0){0.5}}%
\put(837,200){\makebox(0,0){0.25}}%
\put(350,200){\makebox(0,0){0}}%
\put(300,1628){\makebox(0,0)[r]{10}}%
\put(300,1362){\makebox(0,0)[r]{8}}%
\put(300,1097){\makebox(0,0)[r]{6}}%
\put(300,831){\makebox(0,0)[r]{4}}%
\put(300,566){\makebox(0,0)[r]{2}}%
\put(300,300){\makebox(0,0)[r]{0}}%
\end{picture}%
\endgroup
                                      % Figure 3
\caption{Density for the strong standing shock test.   %
The curves are obtained from the three codes, as in    %
Fig.~2.}                                               %
\end{figure}                                           %
%%%%%%%%%%%%%%%%%%%%%%%%%%%%%%%%%%%%%%%%%%%%%%%%%%%%%%%%
The simulations use 100 zones to cover the computational domain, $0 \leq x \leq 1$, and a Courant number of $0.6$. The upstram fluid, in the region $0.5 < x \leq 1$, has density $\rho = 1.0$, pressure $P = 1.0$, and velocity $u = - 1.3\times 10^8$. The downstream fluid, in the region $0 \leq x < 0.5$, has density $\rho = 4$, pressure $P = 1.25\times 10^{16}$, and velocity $u = -3.2\times 10^7$. Ideally the discontinuities should be maintained at $x = 0.5$. Figures 2 and 3 show the pressure and density after $250$ timesteps. The CFL condition for this simulation is determined by the upstream velocity, so the upstream fluid travels a distance of $\sim 0.006$ in each timestep. 

Figure 2 shows that the total energy code produces extremely large errors in the upstream pressure. As expected, Figs.~2 and 3 show that the results from the internal energy code are quite poor. Altogether the hybrid code is the most successful at maintaining the proper pressure and density profiles. For this extreme test, the largest errors from the hybrid code appear in the downstream fluid where the density and pressure differ from their ideal values by at most $11\%$ and $1.6\%$, respectively. The hybrid code also lost about $2.7\%$ of the total energy during the $250$--timestep simulation.

Because the present code conserves linear momentum, there is no erroneous center of mass motion due to numerical errors. 
On the other hand, the code does not conserve angular momentum. This is a fairly serious problem because the angular motion of the star is the central effect we wish to study. The amount of angular momentum that is lost, or gained, through numerical error is found by computing the total angular momentum on the grid, $J_{\rm grid}$, and correcting for the amount of angular momentum that is lost (or gained) through the grid boundaries, $J_{\rm lost}$. The angular momentum passing through the grid boundaries is obtained by computing the angular momentum that moves off the Eulerian grid during each Lagrangian time step. Ideally, the sum of the angular momentum on the grid and the angular momentum lost through the boundaries, $J_{\rm grid} + J_{\rm lost}$, should be constant. As seen in the next section, over the course of the simulation ($8000$ time steps) about 8\% of the star's angular momentum is lost through the boundaries. During this same time, the total angular momentum drops by nearly 25\%. Thus, numerical errors account for a drop of about 17\% in the angular momentum of the star. 

In the Newtonian/quadrupole approximation, the gravitational wave signal $h_{ij}$ is formed from linear combinations of components of the second time derivative of the reduced quadrupole moment,  ${\skew6\ddot{I\mkern-6.8mu\raise0.3ex\hbox{-}}}_{ij}$. The relationship is $h_{ij} = (2G/Rc^4) P_{ij}^{k\ell} {\skew6\ddot{I\mkern-6.8mu\raise0.3ex\hbox{-}}}_{k\ell}$, 
where $P_{ij}^{k\ell}$ is the projection operator for transverse directions and $R$ is the distance from the source to the observation point\cite{MTW}. Using the hydrodynamical equations, the time derivatives that appear in ${\skew6\ddot{I\mkern-6.8mu\raise0.3ex\hbox{-}}}_{ij}$ can be eliminated\cite{BDS} so that, to within boundary terms, we have 
\begin{equation}
   {\skew6\ddot{I\mkern-6.8mu\raise0.3ex\hbox{-}}}_{ij} = 
    {\rm STF}\int d^3x \, 2 \rho
   \left[ v_i v_j - x_{i} \partial_{j} \Phi \right] \ .
\end{equation}
Here, ``STF" stands for the symmetric, trace free part of the expression that follows, $v_i$ is the fluid velocity, $\partial_j = \partial/\partial x_j$ is a spatial derivative, and $\Phi$ is the gravitational potential. 

The gravitational--wave signal is computed numerically by using expression (7) for ${\skew6\ddot{I\mkern-6.8mu\raise0.3ex\hbox{-}}}_{ij}$. 
This calculation is subject to errors due to the finite size of the computational grid. Recall that during the course of the simulation, matter is expelled from the star. Some of this matter, a total of $0.048\,M_\odot$, passes through the grid boundaries. As a consequence, the relationship $h_{ij} = (2G/Rc^4) P_{ij}^{k\ell} {\skew6\ddot{I\mkern-6.8mu\raise0.3ex\hbox{-}}}_{k\ell}$
is not correct, even in the Newtonian/quadrupole approximation, because the lost matter is not included in the calculation of 
${\skew6\ddot{I\mkern-6.8mu\raise0.3ex\hbox{-}}}_{ij}$. 
A second source of error stems from the fact that 
boundary terms were discarded in the derivation of Eq.~(7). These boundary terms would vanish if the density 
were always zero on the grid boundary. Fortunately, the
matter that reaches the boundary has relatively low density and the errors that arise from the missing boundary terms in Eq.~(7) are small. This has been verified by comparing the results from Eq.~(7) with the results obtained by computing numerically the second time 
derivative of the reduced quadrupole moment. The numerical derivatives were obtained by performing a least--squares fit to a quadratic using $5$ or $7$ consecutive data points. The second derivative at the central time is found from the curvature of the quadratic. This calculation is subject to a certain amount of numerical noise, but otherwise the results agree quite closely with those obtained from Eq.~(7). 
Since the matter that passes through the grid boundaries does not 
greatly effect the calculation of ${\skew6\ddot{I\mkern-6.8mu\raise0.3ex\hbox{-}}}_{ij}$ with Eq.~(7), it is plausible to assume that the errors obtained by equating $h_{ij}$ 
and $(2G/Rc^4) P_{ij}^{k\ell} {\skew6\ddot{I\mkern-6.8mu\raise0.3ex\hbox{-}}}_{k\ell}$ 
are also small. 

The frequency spectrum for the gravitational--wave signal is computed with the Lomb normalized periodogram\cite{PTVF}. This method was chosen for its ability to handle the unevenly sampled data. In the Newtonian/quadrupole approximation, the gravitational--wave luminosity is given by $L = G {\skew1\ddot{I\mkern-6.8mu\raise0.3ex\hbox{-}} {\mkern-1.7mu\raise1.9ex\hbox{.}}} \mkern-3mu{}_{ij}
{\skew1\ddot{I\mkern-6.8mu\raise0.3ex\hbox{-}}{\mkern-1.7mu\raise1.9ex\hbox{.}}}\mkern-2mu{}^{ij}/(5c^5)$, and the radiated energy $\Delta E$ is the time integral of $L$\cite{MTW}. The calculation of the time derivative ${\skew1\ddot{I\mkern-6.8mu\raise0.3ex\hbox{-}} {\mkern-1.7mu\raise1.9ex\hbox{.}}} \mkern-3mu{}_{ij} = d{\skew6\ddot{I\mkern-6.8mu\raise0.3ex\hbox{-}}}_{ij}/dt$ 
must be handled with care since, otherwise, even small errors in $L$ will accumulate in the integration over time. For this calculation, I use a Savitzky--Golay approach\cite{PTVF}. Specifically, the time derivative of the function ${\skew6\ddot{I\mkern-6.8mu\raise0.3ex\hbox{-}}}_{ij}$ in each time interval is computed from the slope of a quadratic polynomial that is obtained from a least--squares fit to several nearest neighbor data points. The number of points used is typically 
between $10$ and $20$. 

The evolution code was run on a Cray T90 vector computer, while the initial data, gravitational wave spectrum, and gravitational wave luminosity were computed on local workstations. The evolution required $8000$ timesteps and about $45$ CPU hours. 

%%%%%%%%%%%%%%%%%%%%%%%%%%%%%%%%%%%%%%%%%%%%%%%%%%%%%%%%%%%%%%%%%%%%%%%
\section{Description of the evolution} \label{sec:description}

The star retains its predominantly circular, disk--like shape for about $10\,{\rm ms}$, which amounts to approximately $6$ rotation periods for the fluid near the center of the star. Over the next few milliseconds 
the bar mode grows rapidly, so that by $13\,{\rm ms}$ the star is highly elongated as shown in Fig~4. 
After a few more milliseconds the central regions of the star recircularize, and by $16\,{\rm ms}$ the inner density contours have nearly lost their bar shape as seen in Fig.~5. 
%%%%%%%%%%%%%%%%%%%%%%%%%%%%%%%%%%%%%%%%%%%%%%%%%%%%%
\begin{figure}[!htb]                                %
\begin{picture}(240,250)                            %
\put(0,0){\epsfxsize=3.4in\epsffile{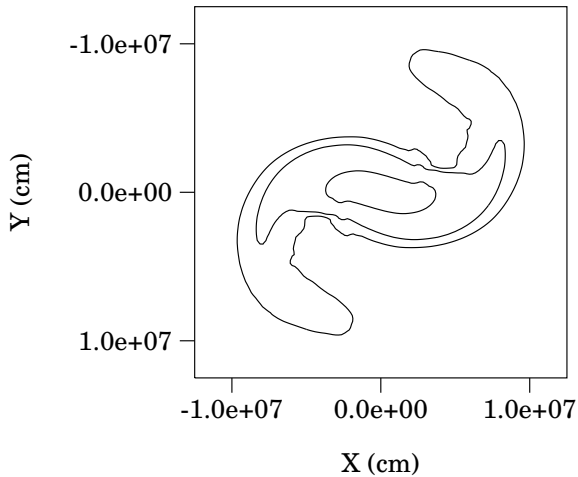}} %
\end{picture}                                       %
\caption{Contour plot of the density in the         %
   equatorial ($X$--$Y$) plane                      % FIGURE 4
   at $13.58\,{\rm ms}$. The contour levels         %
   are $10^{10}\,{\rm g}/{\rm cm}^3$,               %
   $10^{12}\,{\rm g}/{\rm cm}^3$, and               %
   $10^{14}\,{\rm g}/{\rm cm}^3$.}                  %
\end{figure}                                        %
%%%%%%%%%%%%%%%%%%%%%%%%%%%%%%%%%%%%%%%%%%%%%%%%%%%%%
%%%%%%%%%%%%%%%%%%%%%%%%%%%%%%%%%%%%%%%%%%%%%%%%%%%%%
\begin{figure}[!htb]                                %
\begin{picture}(240,250)                            %
\put(0,0){\epsfxsize=3.4in\epsffile{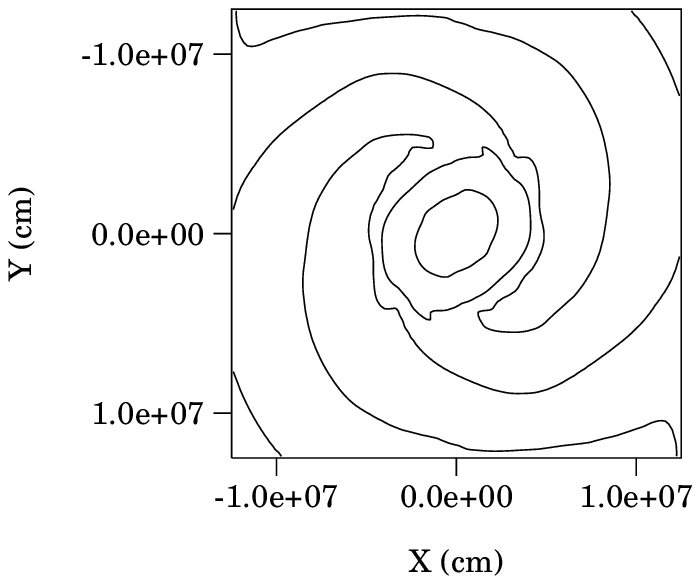}} % FIGURE 5
\end{picture}                                       %
\caption{Density contours as in Fig.~4, at          %
   $16.58\,{\rm ms}$.}                              %
\end{figure}                                        %
%%%%%%%%%%%%%%%%%%%%%%%%%%%%%%%%%%%%%%%%%%%%%%%%%%%%%
%%%%%%%%%%%%%%%%%%%%%%%%%%%%%%%%%%%%%%%%%%%%%%%%%%%%%%
\begin{figure}[!htb]                                 %               
\begin{picture}(240,250)                             %
\put(0,0){\epsfxsize=3.4in\epsffile{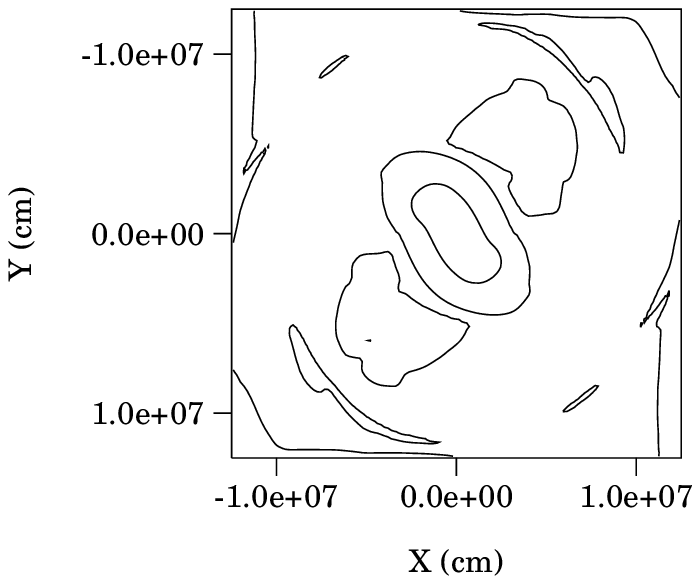}}  % FIGURE 6
\end{picture}                                        %
\caption{Density contours as in Fig.~4,              %
   at $19.48\,{\rm ms}$.}                            %
\end{figure}                                         %
%%%%%%%%%%%%%%%%%%%%%%%%%%%%%%%%%%%%%%%%%%%%%%%%%%%%%%
%%%%%%%%%%%%%%%%%%%%%%%%%%%%%%%%%%%%%%%%%%%%%%%%%%%%%%%%%
\begin{figure}[!htb]                                    %               
\begin{picture}(240,250)                                %
\put(0,0){\epsfxsize=3.4in\epsffile{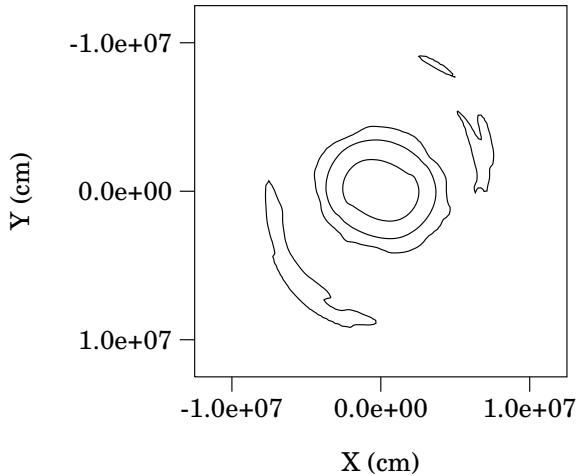}}  % FIGURE 7
\end{picture}                                           %
\caption{Density contours as in Fig.~4, at the end of   %
the simulation ($52.57\,{\rm ms}$).}                    %
\end{figure}                                            %
%%%%%%%%%%%%%%%%%%%%%%%%%%%%%%%%%%%%%%%%%%%%%%%%%%%%%%%%%
The cycle of bar formation and recircularization repeats; by $19\,{\rm ms}$ the star again shows a strong bar--like shape, as shown in Fig.~6. 
The episodes of bar formation and recircularization continue with diminishing amplitude and a period of about $6.5\,{\rm ms}$. During each episode, the bar undergoes a little more than $1\frac{1}{2}$ revolutions. Near the end of the simulation, at a time of $\sim 50\,{\rm ms}$, the oscillations become quite weak and the star settles into a bar--shaped configuration of modest strength. See Fig.~7. 

During the initial episode of rapid bar formation, matter is thrown outward from the ends of the bar. The ejected matter forms long spiral arms, clearly visible in Figs.~4 and 5. Some of this matter is lost off the edge of the computational grid---between $14$ and $20\,{\rm ms}$ the mass loss totals $0.040\,M_\odot$. The subsequent episodes of bar formation are less violent, with only small amounts of ejected mass reaching the grid boundaries. The mass loss between $20\,{\rm ms}$ and the end of the  simulation amounts to $0.008\,M_\odot$. 

When the star assumes its bar shape, the individual particles in the inner regions of the star circulate along trajectories that roughly coincide with the constant density contours. For example, at the time 
shown in Fig.~4, $13.58\,{\rm ms}$, the bar is rotating about the center of mass with a period of approximately $4\,{\rm ms}$. Individual particles flow along the $10^{14}\,{\rm g}/{\rm cm}^3$ density contour with fairly uniform speeds ranging from about $6.5\times 10^9\,{\rm cm}/{\rm s}$ to about $7.5\times 10^9\,{\rm cm}/{\rm s}$. The density contour at $10^{14}\,{\rm g}/{\rm cm}^3$ has a perimeter length of 
approximately $1.7\times 10^7\,{\rm cm}$. Thus, the particles orbit along this contour with period $\sim 2.5\,{\rm ms}$ as the contour precesses about the star's center with period $\sim 4\,{\rm ms}$. 

The shape of the star can be quantified by Fourier analyzing the 
density in a circle in the equatorial plane\cite{TDM,WT}. The Fourier 
coefficients are defined by 
\begin{equation}
   A_m + iB_m = \frac{1}{2\pi} \int_0^{2\pi} d\varphi 
   \, \rho(\varphi) e^{im\varphi} \ ,
\end{equation} 
where $\rho(\varphi)$ is the density in the equatorial plane at an arbitrarily chosen distance of $2.0\times 10^6\,{\rm cm}$ from the center of the star. Here, $\varphi$ is the azimuthal angle. The 
amplitude of the $m^{\rm th}$ Fourier mode is defined by $C_m = \sqrt{A_m^2 + B_m^2}$, and the phase angle is $\phi_m = \arctan(B_m/A_m)$. Figure 8 shows the natural logarithm of the ratio of the amplitude $C_m$ and the average density $C_0$ for the $m=2$ bar mode and the $m=4$ mode.
%%%%%%%%%%%%%%%%%%%%%%%%%%%%%%%%%%%%%%%%%%%%%%%%%%%%
\begin{figure}[!htb]                               %
% GNUPLOT: LaTeX picture with Postscript
\begingroup%
  \makeatletter%
  \newcommand{\GNUPLOTspecial}{%
    \@sanitize\catcode`\%=14\relax\special}%
  \setlength{\unitlength}{0.1bp}%
\begin{picture}(2448,1728)(0,0)%
\special{psfile=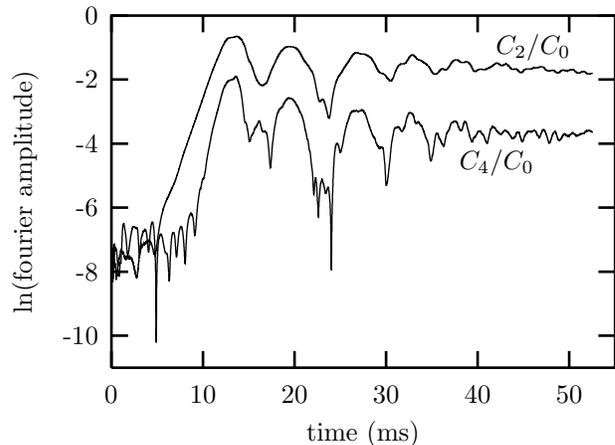 llx=0 lly=0 urx=490 ury=403 rwi=4900}
\put(1711,1073){\makebox(0,0)[l]{$C_4/C_0$}}%
\put(1849,1507){\makebox(0,0)[l]{$C_2/C_0$}}%
\put(1349,50){\makebox(0,0){time (ms)}}%
\put(100,964){%
\special{ps: gsave currentpoint currentpoint translate
270 rotate neg exch neg exch translate}%
\makebox(0,0)[b]{\shortstack{ln(fourier amplitude)}}%
\special{ps: currentpoint grestore moveto}%
}%
\put(2125,200){\makebox(0,0){50}}%
\put(1780,200){\makebox(0,0){40}}%
\put(1435,200){\makebox(0,0){30}}%
\put(1090,200){\makebox(0,0){20}}%
\put(745,200){\makebox(0,0){10}}%
\put(400,200){\makebox(0,0){0}}%
\put(350,1628){\makebox(0,0)[r]{0}}%
\put(350,1387){\makebox(0,0)[r]{-2}}%
\put(350,1145){\makebox(0,0)[r]{-4}}%
\put(350,904){\makebox(0,0)[r]{-6}}%
\put(350,662){\makebox(0,0)[r]{-8}}%
\put(350,421){\makebox(0,0)[r]{-10}}%
\end{picture}%
\endgroup
                                     % Figure 8
\caption{Natural logarithm of the ratios $C_2/C_0$ % 
and $C_4/C_0$ of Fourier amplitudes.}              %
\end{figure}                                       %
%%%%%%%%%%%%%%%%%%%%%%%%%%%%%%%%%%%%%%%%%%%%%%%%%%%%
The $m=3$ mode is small and is not displayed in Fig.~8. The coefficient 
$C_3/C_0$ remains less than $0.01$ through most of the evolution, 
reaching peak values of $\sim 0.02$ in the late stages. 

The periodic growth and decay of the bar shape is clearly seen 
in the peaks and valleys of the graph in Fig.~8. The amplitudes 
grow exponentially for several milliseconds during the initial 
episode of bar formation. For the $m=2$ bar mode, the growth rate near 
$10\,{\rm ms}$ is $d\ln C_2/dt \approx 820/{\rm s}$. For the $m=4$ 
mode, the growth rate is $d\ln C_2/dt \approx 1400/{\rm s}$. In polytropic units ($\tau = 1.88\times 10^{-5}\,{\rm s}$), the growth rates for the $m=2$ and $m=4$ modes are $0.015$ and $0.026$, respectively. For $m=2$, this value differs by a few percent from the result $0.0145$ reported in Ref.\cite{TIPD}. In units of the 
``dynamical time" $t_D = [R_{\rm eq}^3/(GM)]^{1/2}$, where $R_{\rm eq}$ is the initial equatorial radius, the growth rates are $0.50/t_D$ and 
$0.86/t_D$ for the $m=2$ and $m=4$ modes. Comparing these results to the values $0.55/t_D$ and $1.1/t_D$ obtained from the highest resolution run in Ref.~\cite{NCT}, we find a fairly large discrepancy in the $m=4$ case. This difference might be caused by numerical errors in the present code associated with the cartesian grid. 

Near the end of the simulation, the oscillations in the Fourier amplitudes have subsided and the star settles into a bar shape with strength $C_2/C_0 \approx 0.18$. The downward drift in $C_2/C_0$, visible in Fig.~8, is possibly due to numerical error as discussed below. 

The phase angles for the $m=2$ and $m=4$ modes are shown in Fig.~9. 
%%%%%%%%%%%%%%%%%%%%%%%%%%%%%%%%%%%%%%%%%%%%%%%%%
\begin{figure}[!htb]                            %
% GNUPLOT: LaTeX picture with Postscript
\begingroup%
  \makeatletter%
  \newcommand{\GNUPLOTspecial}{%
    \@sanitize\catcode`\%=14\relax\special}%
  \setlength{\unitlength}{0.1bp}%
\begin{picture}(2448,1944)(0,0)%
\special{psfile=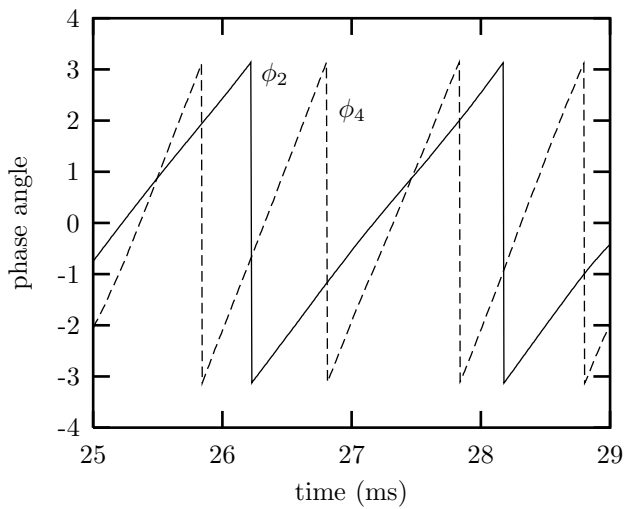 llx=0 lly=0 urx=490 ury=454 rwi=4900}
\put(1275,1497){\makebox(0,0)[l]{$\phi_4$}}%
\put(983,1632){\makebox(0,0)[l]{$\phi_2$}}%
\put(1324,50){\makebox(0,0){time (ms)}}%
\put(100,1072){%
\special{ps: gsave currentpoint currentpoint translate
270 rotate neg exch neg exch translate}%
\makebox(0,0)[b]{\shortstack{phase angle}}%
\special{ps: currentpoint grestore moveto}%
}%
\put(2298,200){\makebox(0,0){29}}%
\put(1811,200){\makebox(0,0){28}}%
\put(1324,200){\makebox(0,0){27}}%
\put(837,200){\makebox(0,0){26}}%
\put(350,200){\makebox(0,0){25}}%
\put(300,1844){\makebox(0,0)[r]{4}}%
\put(300,1651){\makebox(0,0)[r]{3}}%
\put(300,1458){\makebox(0,0)[r]{2}}%
\put(300,1265){\makebox(0,0)[r]{1}}%
\put(300,1072){\makebox(0,0)[r]{0}}%
\put(300,879){\makebox(0,0)[r]{-1}}%
\put(300,686){\makebox(0,0)[r]{-2}}%
\put(300,493){\makebox(0,0)[r]{-3}}%
\put(300,300){\makebox(0,0)[r]{-4}}%
\end{picture}%
\endgroup
                                    % Figure 9
\caption{Phase of the $m=2$ and $m=4$ Fourier   %
   modes.}                                      %
\end{figure}                                    %
%%%%%%%%%%%%%%%%%%%%%%%%%%%%%%%%%%%%%%%%%%%%%%%%%
The eigenfrequencies are $d\phi_2/dt = 3.24/{\rm ms}$ and $d\phi_4/dt = 6.52/{\rm ms}$. The pattern speeds $m^{-1}d\phi_m/dt$ \cite{WT}
are $1.62/{\rm ms}$ for the $m=2$ mode and $1.63/{\rm ms}$ for the $m=4$ mode. These speeds, which equal $0.99/t_D$ and $1.0/t_D$ in dynamical time units, agree with the results obtained in Refs.~\cite{TIPD,NCT} to within $1\%$. The approximate equality of the pattern speeds shows that the $m=4$ mode is a harmonic of the $m=2$ bar mode. 

The stability parameter $\beta$ undergoes fluctuations, dropping to a 
local minimum when the bar amplitude $C_2/C_0$ is a local maximum, and rising to a local maximum when the bar amplitude is a local minimum. This behavior is shown in Fig.~10. 
%%%%%%%%%%%%%%%%%%%%%%%%%%%%%%%%%%%%%%%%%%%%%%%%%
\begin{figure}[!htb]                            %
% GNUPLOT: LaTeX picture with Postscript
\begingroup%
  \makeatletter%
  \newcommand{\GNUPLOTspecial}{%
    \@sanitize\catcode`\%=14\relax\special}%
  \setlength{\unitlength}{0.1bp}%
\begin{picture}(2448,1944)(0,0)%
\special{psfile=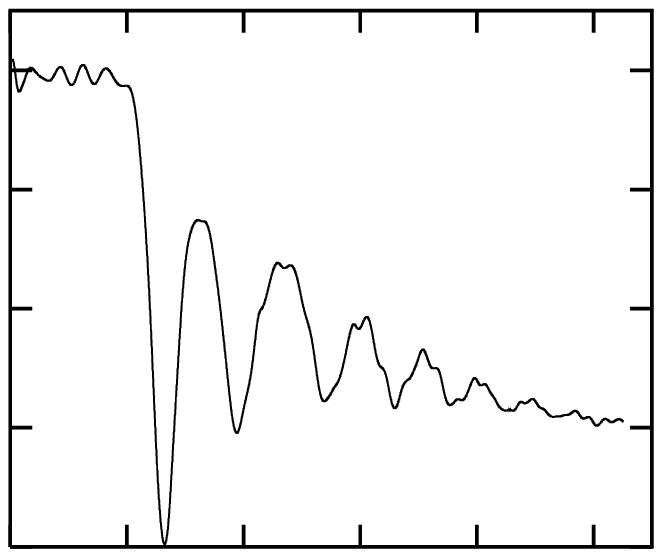 llx=0 lly=0 urx=490 ury=454 rwi=4900}
\put(1374,50){\makebox(0,0){time (ms)}}%
\put(100,1072){%
\special{ps: gsave currentpoint currentpoint translate
270 rotate neg exch neg exch translate}%
\makebox(0,0)[b]{\shortstack{stability parameter}}%
\special{ps: currentpoint grestore moveto}%
}%
\put(2130,200){\makebox(0,0){50}}%
\put(1794,200){\makebox(0,0){40}}%
\put(1458,200){\makebox(0,0){30}}%
\put(1122,200){\makebox(0,0){20}}%
\put(786,200){\makebox(0,0){10}}%
\put(450,200){\makebox(0,0){0}}%
\put(400,1672){\makebox(0,0)[r]{0.3}}%
\put(400,1329){\makebox(0,0)[r]{0.28}}%
\put(400,986){\makebox(0,0)[r]{0.26}}%
\put(400,643){\makebox(0,0)[r]{0.24}}%
\put(400,300){\makebox(0,0)[r]{0.22}}%
\end{picture}%
\endgroup
                                     % Figure 10
\caption{The stability parameter $\beta$ as a   %
   function of time.}                           %
\end{figure}                                    %
%%%%%%%%%%%%%%%%%%%%%%%%%%%%%%%%%%%%%%%%%%%%%%%%%
As the fluctuations diminish, $\beta$ settles to a value of $\sim 0.24$ with a slight downward drift. Note that this value of $\beta$ is well below the critical value $0.27$ for the growth of the bar mode in a constant density star. 

The fact that the bar shape persists with $\beta$ below $0.27$ might indicate that the threshold for bar formation in a $\gamma = 5/3$ star 
with rotation law given by Eq.~(1b) is actually $0.24$ or less. This possibility was tested by evolving a second model star, obtained 
by modifying the freely specifiable data in Eq.~(1) so that the central value of angular velocity is $3000/{\rm s}$ rather than $4000/{\rm s}$. 
The parameters describing this model are 
\begin{mathletters}
\begin{eqnarray}
   M          & = & 1.34\,M_\odot \ ,\\
   R_{\rm eq} & = & 5.88\times 10^6 \,{\rm cm} \ ,\\
   R_{\rm p}  & = & 1.40\times 10^6 \,{\rm cm} \ ,\\
   J          & = & 2.90\times 10^{49}\,{\rm g}\,{\rm cm}^2/{\rm s}\ ,\\
   \beta      & = & 0.253 \ , 
\end{eqnarray}
\end{mathletters}
so, in particular, the stability parameter is between $0.24$ and $0.27$. This model was evolved for approximately $23\,{\rm ms}$, which equals about $11$ rotation periods for the fluid near the center of the star. During this time the $m=2$ bar mode showed no signs of growth. We are thus led to conclude that for a $\gamma = 5/3$ fluid star with rotation law of the type considered here, the dynamical $m=2$ bar mode will not spontaneously grow unless $\beta$ exceeds a critical value near $0.27$; however, once a bar shape is established, it can persist for many rotation periods with $\beta = 0.24$ or less. 

As discussed in the previous section, the numerical code does not conserve angular momentum. In Fig.~11, the total angular momentum on the grid, $J_{\rm grid}$, is plotted as a function of time along with the difference between the initial angular momentum $J_0 = 6.98\times 10^{49}\,{\rm g}\,{\rm cm}^2/{\rm s}$ and the angular momentum $J_{\rm lost}$ that flows off the edge of the numerical grid. 
%%%%%%%%%%%%%%%%%%%%%%%%%%%%%%%%%%%%%%%%%%%%%%%%%
\begin{figure}[!htb]                            %
% GNUPLOT: LaTeX picture with Postscript
\begingroup%
  \makeatletter%
  \newcommand{\GNUPLOTspecial}{%
    \@sanitize\catcode`\%=14\relax\special}%
  \setlength{\unitlength}{0.1bp}%
\begin{picture}(2448,1944)(0,0)%
\special{psfile=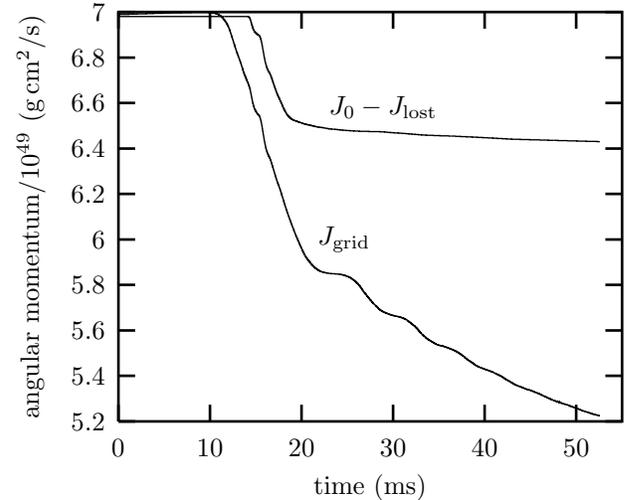 llx=0 lly=0 urx=490 ury=454 rwi=4900}
\put(1194,1484){\makebox(0,0)[l]{$J_0 - J_{\rm lost}$}}%
\put(1152,986){\makebox(0,0)[l]{$J_{\rm grid}$}}%
\put(1349,50){\makebox(0,0){time (ms)}}%
\put(100,1072){%
\special{ps: gsave currentpoint currentpoint translate
270 rotate neg exch neg exch translate}%
\makebox(0,0)[b]{\shortstack{angular momentum/$10^{49}$ 
(${\rm g}\,{\rm cm}^2/{\rm s}$)}}%
\special{ps: currentpoint grestore moveto}%
}%
\put(2125,200){\makebox(0,0){50}}%
\put(1780,200){\makebox(0,0){40}}%
\put(1435,200){\makebox(0,0){30}}%
\put(1090,200){\makebox(0,0){20}}%
\put(745,200){\makebox(0,0){10}}%
\put(400,200){\makebox(0,0){0}}%
\put(350,1844){\makebox(0,0)[r]{7}}%
\put(350,1672){\makebox(0,0)[r]{6.8}}%
\put(350,1501){\makebox(0,0)[r]{6.6}}%
\put(350,1329){\makebox(0,0)[r]{6.4}}%
\put(350,1158){\makebox(0,0)[r]{6.2}}%
\put(350,986){\makebox(0,0)[r]{6}}%
\put(350,815){\makebox(0,0)[r]{5.8}}%
\put(350,643){\makebox(0,0)[r]{5.6}}%
\put(350,472){\makebox(0,0)[r]{5.4}}%
\put(350,300){\makebox(0,0)[r]{5.2}}%
\end{picture}%
\endgroup
                                   %
\caption{The angular momentum on the grid,      %
   $J_{\rm grid}$, and the difference           % Figure 11
   $J_0 - J_{\rm lost}$ between the initial     %
   angular momentum and the angular momentum    %
   lost at the boundaries of the grid.}         %
\end{figure}                                    %
%%%%%%%%%%%%%%%%%%%%%%%%%%%%%%%%%%%%%%%%%%%%%%%%%
Ideally, these two curves should coincide; the fact that $J_{\rm grid}$ falls below $J_0 - J_{\rm lost}$ indicates that numerical errors artificially remove angular momentum from the system. The results for $J_{\rm lost}$ were checked by computing the angular momentum flux from each of the saved data sets, which were generated every 50 timesteps. The numerically computed time derivative of $J_{\rm lost}$ is in very close agreement with the angular momentum flux. 

I have not yet determined the source of the numerical errors in angular momentum. Note, however, that the errors are not significant until the star begins to develop a noticeable bar shape, at about $10\,{\rm ms}$. This suggests that the code has difficulty tracking the leading and trailing edges of the bar--shaped star, with their sharp density gradients, as the bar rotates in the $x$--$y$ plane. 

Since numerical errors cause angular momentum to be lost, it is plausible to expect that rotational kinetic energy is artificially lost as well. This might account for the downward drift in the stability parameter at late times, as seen in Fig.~10. In turn, one might 
expect a drop in the stability parameter to cause the star's bar shape to decay. Figure 8 shows that the Fourier amplitude $C_2$ is fairly robust at late times with a slight downward trend on average. If angular momentum were not artificially lost, the simulation might show that after the fluctuations cease the bar shape remains unchanged on a dynamical time scale. 

%%%%%%%%%%%%%%%%%%%%%%%%%%%%%%%%%%%%%%%%%%%%%%%%%%%%%%%%%%%%%%%%%%%%%%%
\section{Gravitational waves} \label{sec:waves}

In the Newtonian/quadrupole approximation, the gravitational wave amplitude for the plus polarization state, $h_+$, as measured at a distance $R$ in the equatorial plane of the source, is 
\begin{eqnarray}
   \frac{c^4 R}{G} h_+({\rm eq}) & = & (\cos^2\varphi - 2)
     {\skew6\ddot{I\mkern-6.8mu\raise0.3ex\hbox{-}}}_{xx}
       + (\sin^2\varphi - 2)         {\skew6\ddot{I\mkern-6.8mu\raise0.3ex\hbox{-}}}_{yy} \nonumber\\
      & & + 2\cos\varphi \sin\varphi \,
        {\skew6\ddot{I\mkern-6.8mu\raise0.3ex\hbox{-}}}_{xy} \ .
\end{eqnarray}
Here,  ${\skew6\ddot{I\mkern-6.8mu\raise0.3ex\hbox{-}}}_{ij}$ is 
the second time derivative of the reduced quadrupole moment and $\varphi$ is the azimuthal angle of the observation point relative to the (arbitrary) $x$--axis of the source. Figure 12 shows the results for $h_+$ with $R = 20\,{\rm Mpc}$ and $\varphi = \pi/2$. 
%%%%%%%%%%%%%%%%%%%%%%%%%%%%%%%%%%%%%%%%%%%%%%%%%
\begin{figure}[!htb]                            %
% GNUPLOT: LaTeX picture with Postscript
\begingroup%
  \makeatletter%
  \newcommand{\GNUPLOTspecial}{%
    \@sanitize\catcode`\%=14\relax\special}%
  \setlength{\unitlength}{0.1bp}%
\begin{picture}(2448,1728)(0,0)%
\special{psfile=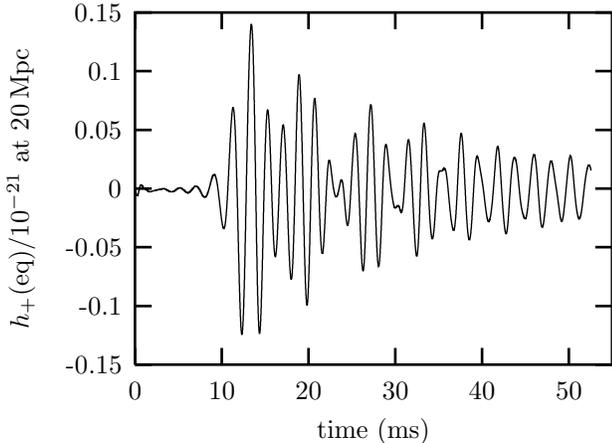 llx=0 lly=0 urx=490 ury=403 rwi=4900}
\put(1399,50){\makebox(0,0){time (ms)}}%
\put(100,964){%
\special{ps: gsave currentpoint currentpoint translate
270 rotate neg exch neg exch translate}%
\makebox(0,0)[b]{\shortstack{$h_+({\rm eq})/10^{-21}$ at $20\,{\rm Mpc}$}}%
\special{ps: currentpoint grestore moveto}%
}%
\put(2135,200){\makebox(0,0){50}}%
\put(1808,200){\makebox(0,0){40}}%
\put(1481,200){\makebox(0,0){30}}%
\put(1154,200){\makebox(0,0){20}}%
\put(827,200){\makebox(0,0){10}}%
\put(500,200){\makebox(0,0){0}}%
\put(450,1628){\makebox(0,0)[r]{0.15}}%
\put(450,1407){\makebox(0,0)[r]{0.1}}%
\put(450,1185){\makebox(0,0)[r]{0.05}}%
\put(450,964){\makebox(0,0)[r]{0}}%
\put(450,743){\makebox(0,0)[r]{-0.05}}%
\put(450,521){\makebox(0,0)[r]{-0.1}}%
\put(450,300){\makebox(0,0)[r]{-0.15}}%
\end{picture}%
\endgroup
                                  %
\caption{The gravitational--wave amplitude      %
   $h_+({\rm eq})$ measured at an observation   % FIGURE 12
   point in the equatorial plane of the         %
   source at a distance of $20\,{\rm Mpc}$.}    %
\end{figure}                                    %
%%%%%%%%%%%%%%%%%%%%%%%%%%%%%%%%%%%%%%%%%%%%%%%%%
(Recent estimates indicate that the distance to the Virgo cluster 
is $\sim 20.9\,{\rm Mpc}$\cite{TSR}.) In the equatorial plane, the cross polarization amplitude, $h_\times({\rm eq})$, is a linear combination of the components 
${\skew6\ddot{I\mkern-6.8mu\raise0.3ex\hbox{-}}}_{xz}$ and 
${\skew6\ddot{I\mkern-6.8mu\raise0.3ex\hbox{-}}}_{yz}$. Since the star is approximately symmetric under reflections in the equatorial plane, $h_\times({\rm eq})$ as measured in the equatorial plane is nearly zero. 

For an observation point above the north pole of the star, the plus polarization amplitude is 
\begin{eqnarray}
   \frac{c^4 R}{G} h_+({\rm p}) & = & (2\cos^2\varphi - 1)
     {\skew6\ddot{I\mkern-6.8mu\raise0.3ex\hbox{-}}}_{xx}
       + (2\sin^2\varphi - 1) {\skew6\ddot{I\mkern-6.8mu\raise0.3ex\hbox{-}}}_{yy} \nonumber\\
      & & + 4\cos\varphi \sin\varphi \,
        {\skew6\ddot{I\mkern-6.8mu\raise0.3ex\hbox{-}}}_{xy} \ .
\end{eqnarray}
Since the star is highly flattened and has little motion in the $z$--direction, the $zz$ component of the second time derivative of the {\em unreduced} quadrupole moment, ${\ddot I}_{zz}$, is very small. Furthermore, because the shape of the bar does not evolve rapidly compared to its rotation rate, the $xx$ and $yy$ components satisfy 
${\ddot I}_{xx} + {\ddot I}_{yy} \approx 0$. These observations imply  
${\skew6\ddot{I\mkern-6.8mu\raise0.3ex\hbox{-}}}_{xx} \approx - 
{\skew6\ddot{I\mkern-6.8mu\raise0.3ex\hbox{-}}}_{yy}$. As a consequence, we see from Eqs.~(10) and (11) that $h_+({\rm p}) \approx 2 h_+({\rm eq})$. This is confirmed by the results of the simulation. As expected, the amplitude $h_\times({\rm p})$ is almost identical to $h_+({\rm p})$, just phase shifted by $45^\circ$. 

Although the reliability of the results for the gravitational--wave signal is rather poor, due both to the crudeness of the physical model and to numerical errors, it is still of some interest to consider the detectability of these waves. Thus, consider the characteristic amplitude of the source, approximated by\cite{Thorne}
\begin{equation}
   h_c \approx \left( \frac{3G\Delta E}{2\pi^2 c^3 f_c R^2} \right)^{1/2} \ .
\end{equation}
Here, $\Delta E$ is the total energy radiated in gravitational waves 
and $f_c$ is the characteristic frequency. The gravitational--wave frequency is sharply peaked about $490\,{\rm Hz}$, as shown in Fig.~13. 
%%%%%%%%%%%%%%%%%%%%%%%%%%%%%%%%%%%%%%%%%%%%%%%%%
\begin{figure}[!htb]                            %
% GNUPLOT: LaTeX picture with Postscript
\begingroup%
  \makeatletter%
  \newcommand{\GNUPLOTspecial}{%
    \@sanitize\catcode`\%=14\relax\special}%
  \setlength{\unitlength}{0.1bp}%
\begin{picture}(2448,1728)(0,0)%
\special{psfile=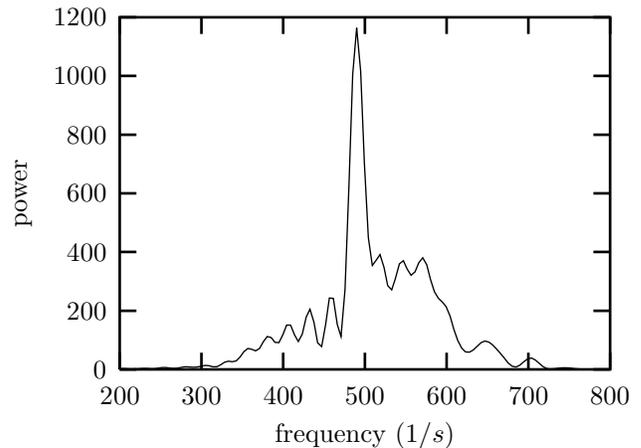 llx=0 lly=0 urx=490 ury=403 rwi=4900}
\put(1374,50){\makebox(0,0){frequency ($1/s$)}}%
\put(100,964){%
\special{ps: gsave currentpoint currentpoint translate
270 rotate neg exch neg exch translate}%
\makebox(0,0)[b]{\shortstack{power}}%
\special{ps: currentpoint grestore moveto}%
}%
\put(2298,200){\makebox(0,0){800}}%
\put(1990,200){\makebox(0,0){700}}%
\put(1682,200){\makebox(0,0){600}}%
\put(1374,200){\makebox(0,0){500}}%
\put(1066,200){\makebox(0,0){400}}%
\put(758,200){\makebox(0,0){300}}%
\put(450,200){\makebox(0,0){200}}%
\put(400,1628){\makebox(0,0)[r]{1200}}%
\put(400,1407){\makebox(0,0)[r]{1000}}%
\put(400,1185){\makebox(0,0)[r]{800}}%
\put(400,964){\makebox(0,0)[r]{600}}%
\put(400,743){\makebox(0,0)[r]{400}}%
\put(400,521){\makebox(0,0)[r]{200}}%
\put(400,300){\makebox(0,0)[r]{0}}%
\end{picture}%
\endgroup
                                % FIGURE 13
\caption{The frequency spectrum of              %
   gravitational waves, computed with the       %
   Lomb normalized periodogram.}                %
\end{figure}                                    %
%%%%%%%%%%%%%%%%%%%%%%%%%%%%%%%%%%%%%%%%%%%%%%%%%
The total energy radiated in gravitational waves, as computed in the 
Newtonian/quadrupole approximation up to the time the simulation
was halted, is $\Delta E \approx 5\times 10^{-4}\,{\rm M}_\odot {\rm c}^2$. From these data we find the characteristic amplitude of the source to be $h_c \approx 4\times 10^{-22}$ at a distance of $20\,{\rm Mpc}$. This value for $h_c$ is a lower bound, since the radiated energy 
$\Delta E$ will continue to increase as long as the bar persists. Also 
note that the result $h_c \approx h\sqrt{n} \approx 4\times 10^{-22}$ coincides roughly with $n\approx 25$ wave cycles with amplitude $h\approx 0.08\times 10^{-21}$. This is consistent with the signal displayed in Fig.~12. 

The signal $h_c \approx 4\times 10^{-22}$ is probably too weak to be seen by the first generation of interferometric gravitational--wave detectors, which should have a sensitivity of about $10^{-20}$ at $500\,{\rm Hz}$\cite{BandC,Thorne2}. The signal might be detectable by the planned advanced detectors, which are expected to have a sensitivity below $10^{-21}$ at $500\,{\rm Hz}$\cite{BandC,Thorne2}. For a source at a distance of $\sim 25\,{\rm kpc}$, comparable to the diameter of the Milky Way galaxy, the characteristic amplitude is $h_c \approx 3\times 10^{-19}$. This is a relatively large signal that should be detected easily by the first generation of interferometers. 

If the simulation is scaled as described in section II, the amplitude as measured at a fixed (unscaled) distance $R$ changes by a factor of $\tau^2/(\mu\lambda^2)$ and the frequency changes by a factor of $\tau$. For example, with $\lambda = 1/2$, $\mu = 2$, and $\tau = 1/4$, the amplitude $h_+({\rm eq})$ at $R = 20\,{\rm Mpc}$ is reduced from that shown in Fig.~12 by a factor of $8$. The characteristic amplitude $h_c$ is also reduced by a factor of $8$, to about $5\times 10^{-23}$ at $20\,{\rm Mpc}$, and the peak frequency occurs at $\sim 120\,{\rm Hz}$. This signal is somewhat below the expected sensitivity of the advanced gravitational--wave detectors, which is approximately $10^{-22}$ at $120\,{\rm Hz}$\cite{BandC,Thorne2}. At a distance of $25\,{\rm kpc}$, the characteristic amplitude is almost $4\times 10^{-20}$, well above the expected sensitivity of $\sim 4\times 10^{-21}$ for the first generation detectors.

%%%%%%%%%%%%%%%%%%%%%%%%%%%%%%%%%%%%%%%%%%%%%%%%%%%%%%%%%%%%%%%%%%%%%%%
\section{Conclusions} \label{sec:conclusions}
The simulation presented here shows that a rapidly rotating $\gamma = 5/3$ fluid star with rotation law Eq.~(1b) is dynamically unstable to bar formation, and that the bar shape, once established, is relatively long lived. Over the final $10$ or $20\,{\rm ms}$ of the simulation, the star maintains a modest but fairly persistent bar shape with stability parameter $\beta$ equal to about $0.24$. It is possible that, for $\gamma = 5/3$ stars with certain initial angular velocity profiles, the threshold for bar formation is actually $0.24$ or less. However, further numerical study showed that an axisymmetric star with $\beta = 0.25$ 
initially is stable against growth of the bar mode. Thus, for the models considered here, the dynamical bar mode requires a value of $\beta$ near $0.27$ to grow but the bar shape can persist for many bar--rotation periods with $\beta = 0.24$ or less. Imamura, Durisen, and Pickett suggest that as long as $\beta$ is greater than the critical value for growth of the secular instability, about $0.14$, a bar--shaped star is dynamically stable against forming an axisymmetric disk\cite{IDP}. The results here are consistent with this conjecture. Recent numerical studies by New, Centrella, and Tohline\cite{NCT} also indicate that the bar shape is long lived. 

The strength of the gravitational--wave signal emitted by an initially 
axisymmetric, dynamically unstable star depends on the rotating star's length and time scales. With the numerical model scaled to represent a low--density neutron star with soft equation of state, the characteristic amplitude is $h_c \approx 4\times 10^{-22}$ at 
a distance of $20\,{\rm Mpc}$. With the model scaled to represent 
a stellar core that has partially collapsed and is prevented from contracting to nuclear densities by centrifugal forces, the characteristic amplitude is $h_c \approx 5\times 10^{-23}$ at $20\,{\rm Mpc}$. These signals are probably too weak to be detected by the first generation of interferometers, although the neutron star signal might be detectable by the planned advanced interferometers. At a distance of $25\,{\rm kpc}$, both signals are quite large and should be detected easily by the first generation interferometers. 

%%%%%%%%%%%%%%%%%%%%%%%%%%%%%%%%%%%%%%%%%%%%%%%%%%%%%%%%%%%%%%%%%%%%%%%
\section*{Acknowledgments}\label{sec:ack}
I would like to thank John Blondin for helpful discussions. Numerical calculations were carried out on the Cray T90 at the North Carolina Supercomputing Center. 

%%%%%%%%%%%%%%%%%%%%%%%%%%%%%%%%%%%%%%%%%%%%%%%%%%%%%%%%%%%%%%%%%%%%%%%
\onecolumn
\section*{Appendix}
The technique used in this work to solve the equations of hydrostatic equilibrium is similar to the self--consistent field method, first developed by Ostriker and Mark\cite{OM} and later refined by Hachisu\cite{Hachisu}. The self--consistent field method is based on the integral form of the Euler equation which reads, for the equation of state $P = K\rho^\gamma$\cite{OM,Hachisu}, 
\begin{equation}
   \frac{\gamma K}{\gamma - 1} \rho^{\gamma-1} + \Psi + \Phi = C \ .
\end{equation}
In this equation, $\Psi = -\int dr\, r\, \omega^2$ is the ``rotational potential" derived from the angular velocity $\omega(r)$, $\Phi$ is the gravitational potential, and $C$ is an integration constant. In the self--consistent field method, one begins with an initial guess for the density $\rho$, solves the Poisson equation for $\Phi$, then uses Eq.~(13) to obtain a corrected density distribution. The process is then iterated. Typically, the Poisson equation is solved by expanding the potential $\Phi$ in terms of Legendre polynomials, and the constant $C$ appearing in Eq.~(13) is determined by specifying certain properties at the star's surface. For example, Hachisu fixes the ``axis ratio" $R_{\rm p}/R_{\rm eq}$\cite{Hachisu}. 

For the method used in this paper, the constant $C$ is determined by properties at the center of the star. The rotational potential is chosen to vanish at the star's center, $\Psi(0) = 0$, and the constant $C$ is written as 
\begin{equation}
   C = \gamma K \rho_{\rm c}^{\gamma-1} + \Phi_{\rm c} \ .
\end{equation}
Here, the subscript c denotes the center of the star. The central density $\rho_{\rm c}$ is specified as input data, along with the rotation law $\omega(r)$. The initial guess for $\rho$ is chosen to be a static, spherically symmetric polytrope with central density $\rho_{\rm c}$, obtained by solving numerically the Lane--Emden equation. The potential $\Phi$, and its central value, are obtained from the Poisson equation using a multigrid algorithm, as discussed below. Note that with this scheme the center of the star must have nonzero density. This precludes the possibility of generating stars with toroidal surfaces. 

Equation (13) can be written as $F=0$ where the function $F$ is defined by 
\begin{equation}
   F \equiv \frac{\gamma K}{\gamma - 1} \rho^{\gamma-1} + \Psi 
   + \Phi - C 
\end{equation}
and the constant $C$ is given by Eq.~(14). The corrections to the density are obtained from a Newton--Raphson algorithm applied to $F=0$. Note that $F(x)$ is a nonlocal function of $\rho(x)$, where $x$ labels  points in space. The nonlocal dependence of $F(x)$ on $\rho(x)$ enters through the gravitational potential $\Phi(x)$. One might expect that the value of $\Phi$ at the point $x$ is insensitive to the density $\rho(x)$ at the same point $x$, since the potential $\Phi$ is determined primarily by the global properties of the star, such as its mass. With this observation in mind, we compute the variation of $F$ at point $x$ with respect to $\rho$ at point $x$ by dropping the nonlocal terms. This leads to the approximate result $\delta F \approx \gamma K \rho^{\gamma-2} \delta \rho$. Then setting $F + \delta F = 0$, we find the density correction
\begin{equation}
   \delta \rho = \frac{1}{\gamma K} \rho^{2-\gamma} (C - \Phi - \Psi) -     \frac{1}{\gamma - 1} \rho \ ,
\end{equation}
to be used for each Newton--Raphson iteration. 

I have not constructed a mathematical argument to justify the above assumption, namely, that the nonlocal terms in $F$ can be ignored in computing $\delta F$. However, this same assumption is implicit in the usual self--consistent field method. As described in Refs.~\cite{OM,Hachisu}, for the self--consistent field method the density is updated by solving Eq.~(13) for $\rho$, ignoring the $\rho$--dependence in $\Phi$. That is, the change in $\rho$ is 
\begin{equation}
   \delta \rho = \left[ \frac{\gamma - 1}{\gamma K} (C - \Phi - \Psi)       \right]^{\frac{1}{\gamma-1}} - \rho \ ,
\end{equation}
where $\Phi$ is the solution of the Poisson equation. Formula (17) can be justified by the same reasoning that led to Eq.~(16), but with the following difference: enthalpy $H \equiv \gamma K \rho^{\gamma-1}/(\gamma - 1)$, rather than density $\rho$, is treated as the independent variable. The variation of $F$ with respect to $H$, ignoring the $\Phi$ dependence on $H$, is $\delta F \approx \delta H$. Setting $F + \delta F = 0$ leads to a new enthalpy distribution, $H_{\rm new} = H - F$. Solving this equation for the new density $\rho_{\rm new}$, we find the density correction $\delta\rho = \rho_{\rm new} - \rho$ of Eq.~(17). 

The initial data code solves the equations of hydrostatic equilibrium at the zone centers of a $513\times 513$ grid in the $r$--$z$ plane of a cylindrical coordinate system. The symmetry axis concides with the edge of the first column of zones. 
The density $\rho$ is computed on the inner $512\times 511$ zones. One 
layer of zones at the top, bottom, and side opposite the symmetry axis are reserved for setting boundary conditions for the gravitational potential $\Phi$. The boundary values are obtained by expanding the potential in a  multipole series that includes monopole and quadrupole terms. (The dipole and octupole moments vanish due to rotational symmetry and reflection symmery about the equatorial plane.) The Poisson equation is solved with a multigrid algorithm, adapted from the code described in Ref.~\cite{PTVF} to handle the boundary conditions and the cylindrical geometry. 

The initial data described in Sec.~II was generated by a code that relies on the Newton--Raphson iteration scheme (16) and the multigrid Poisson solver described above. The results of three tests of this code (JDB) are shown in Table I. For comparison, the results obtained by Hachisu\cite{Hachisu} are also listed. The values reported in the table are scaled as in Hachisu's self--consistent field method, with Newton's constant $G$, the equatorial radius $R_{\rm eq}$, and the maximum density $\rho_{\rm max}$ equal to unity. Hats denote these scaled quantities. Also, $\Pi$ denotes the volume integral of pressure, $\int P\,dV$. 
\setlength{\tabcolsep}{8pt}
\begin{center}
\begin{tabular}{r|c|lllllllc}
& test & $\ {\hat R}_{\rm p}$ & $\ {\hat M}$ & $\ {\hat J}$ & $\ {\hat T}$ & $-{\hat W}$ & $\ 3{\hat\Pi}$ & ${\hat P}_{\rm max}$ & ${\cal V}$ \\ \hline
JDB & $r$ & $0.6667$ & $0.3288$ & $0.02575$ & $0.006641$ & $0.1164$ & $0.1031$ & $0.2044$ & $1.8\times 10^{-4}$ \\
Hachisu & $r$ & $0.667$ & $0.328$ & $0.0257$ & $0.00663$ & $0.116$ & $0.103$ & $0.204$ & \\ [1.0ex]
JDB & $v$ & $0.3332$ & $0.6413$ & $0.1378$ & $0.06392$ & $0.3733$ & $0.2454$ & $0.2020$ & $4.7\times 10^{-5}$ \\
Hachisu & $v$ & $0.333$ & $0.639$ & $0.137$ & $0.0638$ & $0.372$ & $0.244$ & $0.202$ & \\ [1.0ex]
JDB & $j$ & $0.1662$ & $0.8419$ & $0.1036$ & $0.04559$ & $0.5982$ & $0.5070$ & $0.3272$ & $1.6\times 10^{-5}$ \\
Hachisu & $j$ & $0.167$ & $0.843$ & $0.104$ & $0.0456$ & $0.599$ & $0.508$ & $0.327$ &
\end{tabular}
\end{center}
The ``$r$" test assumes a rigid rotation law, $\omega(r) = \omega_0$, with the scaled angular velocity given by ${\hat\omega}_0^2 = 0.266$. The ``$v$" test assumes the ``$v$--constant" rotation law, $\omega(r) = v_0/\sqrt{d^2 + r^2}$, with ${\hat v}_0^2 = 0.215$ and ${\hat d} = 0.100$. The ``$j$" test assumes the ``$j$--constant" rotation law, $\omega(r) = j_0/(d^2 + r^2)$, with ${\hat j}_0^2 = 0.0176$ and ${\hat d} = 0.100$. For each test, $\gamma = 5/3$. For Hachisu's data the virial theorem diagnostic ${\cal V}$ is typically less than a few times $10^{-4}$\cite{Hachisu}. The test results show that the present code and Hachisu's code differ by less than $1\%$ in every case.

Subsequent testing has shown that the density update in Eq.~(16) is no better, in terms of accuracy and convergence rate, than the usual self--consistent field method update (17). After $30$ Newton--Raphson iterations, the results obtained from the two update formulas agree to a few parts in $10^{-9}$, at worst. For both update formulas, the errors (as compared to the $30$--iteration results) were at most a few tenths of a percent after $10$ iterations. 

%%%%%%%%%%%%%%%%%%%%%%%%%%%%%%%%%%%%%%%%%%%%%%%%%%%%%%%%%%%%%%%%%%%%%%%
\twocolumn

%%%%%%%%%%%%%%%%%%%%%%%%%%%%%%%%%%%%%%%%%%%%%%%%%%%%%%%%%%%%%%%%%%%%%%%
\end{document}